\documentclass[twocolumn,showpacs,aps,prl]{revtex4}

\usepackage{graphicx}
\usepackage{dcolumn}
\usepackage{amsmath}
\usepackage{epsfig}
\long\def\inst#1{\par\nobreak\kern 4pt\nobreak
    {\itshape #1}\par\vskip 10pt plus 3pt minus 3pt}
\RequirePackage{xspace}
\usepackage{relsize}

\begin{document}

\title{
\large \bfseries \boldmath A Weighting Method for Incorporating Mass Resolution Effects in Amplitude Analysis}

\author{Benhou Xiang}\email{xiangbh@ihep.ac.cn}

\affiliation{ University of Chinese Academy of Sciences,
                  Beijing  100049, China}
\affiliation{  Institute of High Energy Physics,
                 Chinese Academy of Sciences, Beijing  100049, China}

\author{Wenqian Zheng}\email{zhengwq@ihep.ac.cn}
\affiliation{China University of Geosciences, Wuhan 430074, China}
\affiliation{  Institute of High Energy Physics,
                  Chinese Academy of Sciences, Beijing  100049, China}
\author{Hongxun Yang}\email{yanghx@ihep.ac.cn}
\affiliation{  Institute of High Energy Physics,
                  Chinese Academy of Sciences, Beijing  100049, China}
\author{Xiaolin Kang}\email{kangxiaolin@cug.edu.cn}
\affiliation{China University of Geosciences, Wuhan 430074,  China}

\author{Shuangshi Fang}\email{fangss@ihep.ac.cn}
\affiliation{  Institute of High Energy Physics,
                  Chinese Academy of Sciences, Beijing  100049, China}
\affiliation{ University of Chinese Academy of Sciences,
                  Beijing  100049, China}
\affiliation{   School of Physics, Henan Normal University, Xinxiang 453007, China
                  }


\date{\today}


\begin{abstract}

We present a phase-space weighting method for incorporating detector mass-resolution effects into amplitude analysis in an efficient and flexible way. The method uses fully simulated Monte Carlo samples containing both truth-level and reconstructed kinematics to estimate the local detector response around each observed event. Based on these truth–reconstruction correspondences, event-by-event weights are constructed to map theoretical amplitudes from the true phase space to the smeared observational space. In this way, detector smearing can be included in the likelihood fit without performing an explicit multidimensional convolution or on-the-fly detector simulation. The procedure naturally handles nonuniform resolution across the Dalitz plane and can be extended beyond invariant masses to momentum and angular variables. Applied to the decays $J/\psi \rightarrow \Xi^{-} \bar{\Xi}^{+}\pi^0$  and $J/\psi\rightarrow K^-\Lambda\bar{\Xi}^+$, the method significantly reduces biases in fitted resonance parameters and interference patterns, suppresses artificial  structures, and improves the reliability of partial-wave analysis and resonance spectroscopy in high-precision hadron experiments.
 
\end{abstract}


\maketitle

\section{Introduction}
Amplitude analysis, or Partial Wave Analysis (PWA), is a crucial technique in high energy physics for extracting the properties of resonant states produced in particle collisions. It involves fitting a theoretical model, which describes the production and decay dynamics via a coherent sum of partial waves, to the observed distribution of final state particles. The accuracy of this technique hinges on the precise correspondence between the theoretical model, which describes the ``true'' underlying physics, and the experimentally ``observed'' data.

A fundamental challenge in this process arises from the finite resolution of particle detectors. The measured momenta of final state tracks inevitably deviate from their true values due to detector effects such as material interactions, spatial resolution, and reconstruction algorithms. This smearing distorts the observed invariant mass distributions and other kinematic observables, particularly for narrow resonances. The theoretical model, which is formulated for the true phase space, cannot accurately describe these resolution-smeared distributions. Consequently, ignoring resolution effects can lead to significant systematic biases: the measured masses and widths of resonances may be inaccurate, the determination of spin-parity quantum numbers can be compromised, and, most critically, statistically significant ``fake'' resonant structures may appear in the analysis to artificially improve the description of the deformed line shapes. This issue is especially pertinent for analyses involving narrow states like $\omega$, $\phi(1020)$, or potential new particles, where the resonance width is comparable to or smaller than the experimental resolution.

Traditional methods to account for resolution, such as performing a full detector simulation and reconstruction for each point in the parameter space during the fit, are computationally prohibitive. They become even more intractable for multi-dimensional resolution effects or for processes with complex final states. Therefore, there is a pressing need for an efficient, robust, and computationally lightweight technique to incorporate resolution effects directly into the PWA framework.

This article introduces a phase-space weighting method designed to address this challenge. The key idea is to construct a set of weights that map the theoretical predictions from the ``true'' phase space to the ``smeared'' or ``detected'' phase space. This method is not only highly efficient but also capable of handling mass distribution resolutions without a significant increase in computational cost. We will describe the methodology, demonstrate its application and performance using the decay \(J/\psi \rightarrow \Xi^{-} \bar{\Xi}^{+}\pi^{0}\) as a concrete example, and discuss its implications for extending the scope and reliability of amplitude analyses in high energy physics experiments.

\section{Notation and Method Overview}
\label{sec:weighting_method}
The goal of a specific analysis is to infer the true value of an observable from experimentally detected events, which are distorted by the detector response. In particular, we aim to recover the truth-level differential cross section from the reconstructed one.
However,   in amplitude analyses of multi-body decays, the observed event distribution is distorted by the finite detector resolution. 
This effect is particularly important for narrow intermediate states, for which mass smearing can significantly deform the line shape and bias the extracted resonance parameters.
A full treatment of detector resolution would require the convolution of the theoretical intensity with the detector response function over the multidimensional phase space. 
In practice, such a treatment often requires too much time and computational resources for an amplitude analysis.

To address this issue, we introduce an event-by-event weighting method that approximates the local deconvolution of the detector response. 
Since the true distribution is not available in real data, we illustrate the method with a numerical study based on the fully simulated Monte Carlo (MC) events for a specific process.
The basic idea is to assign an event-by-event resolution-correction weight that locally approximates the deconvolution of the detector response. This is achieved by using nearby MC events for which both truth-level and reconstructed kinematics are known. The procedure is iterative and can be naturally incorporated into an amplitude analysis.


The basic idea is to construct, for each reconstructed data event, a local estimate of the detector smearing from nearby MC events for which both the truth-level and reconstructed kinematics are known. This estimate is then used to define a resolution-correction factor that relates the observed local event density to the underlying truth-level intensity predicted by the amplitude .

Consider a reconstructed data event measured at phase-space point $x_0$. Around this point, we select a set of nearby MC events that contain both truth and reconstructed kinematics. For each selected MC event $i=1,\ldots,n$, we define the truth--reconstruction shift for the relevant kinematic variables, for example the invariant-mass-squared variables. 

Let $x_0$ denote the reconstructed phase-space coordinates of a given data event.
For a three-body decay, $x_0$ may be represented by two independent invariant-mass-squared variables in the Dalitz plane, while the third one is kinematically constrained.
To characterize the local detector response around $x_0$, we select a set of nearby fully simulated MC events in reconstructed phase space.
For each selected MC event $i=1,\dots,n$, we define the truth--reconstruction shift vector
\begin{equation}
\Delta x_i = x_{i,\mathrm{truth}} - x_{i,\mathrm{reco}},
\end{equation}
where $x_{i,\mathrm{truth}}$ and $x_{i,\mathrm{reco}}$ are the truth-level and reconstructed phase-space coordinates of the $i$th MC event, respectively.
 The vector $\Delta x_i$ therefore describes how detector smearing shifts the reconstructed event relative to its truth value.

The collection $\{\Delta x_i\}$ obtained from MC events in the neighborhood of $x_0$ provides a local empirical description of the detector response, including both resolution effects and possible reconstruction biases.

Given an amplitude model $A(x)$, we evaluate the model at the measured point. 
Let $A(x)$ denote the complex decay amplitude at phase-space point $x$, and let
\begin{equation}
I(x) \equiv |A(x)|^2
\end{equation}
be the corresponding theoretical intensity.
For the reconstructed data event at $x_0$, we evaluate the model intensity at the measured point,
\begin{equation}
I(x_0) = |A(x_0)|^2,
\end{equation}
and at the locally shifted points
\begin{equation}
x_i = x_0 + \Delta x_i,
\qquad i=1,\dots,n,
\end{equation}
which approximate possible truth-level phase-space points that could migrate to $x_0$ after detector smearing.

The locally smeared intensity is then estimated by averaging over the nearby MC-induced shifts,
\begin{equation}
\overline{I}(x_0) = \frac{1}{n}\sum_{i=1}^{n} I(x_0+\Delta x_i)
= \frac{1}{n}\sum_{i=1}^{n} |A(x_0+\Delta x_i)|^2.
\label{eq:avg_intensity}
\end{equation}
In practice, the number of nearby MC events is chosen sufficiently large, typically $n \gtrsim 100$, to suppress statistical fluctuations while maintaining locality in phase space.

We then define the event-by-event resolution-correction factor as
\begin{equation}
f_{\mathrm{corr}}(x_0) = \frac{I(x_0)}{\overline{I}(x_0)}
= \frac{|A(x_0)|^2}{\frac{1}{n}\sum_{i=1}^{n}|A(x_0+\Delta x_i)|^2}.
\label{eq:f_reso}
\end{equation}
This factor quantifies the ratio of the truth-level model intensity at $x_0$ to the locally smeared intensity expected at the same reconstructed point.
For a flat phase-space distribution, detector smearing redistributes events without altering the local average intensity, apart from efficiency and acceptance effects.
In contrast, when resonant structures are present, the local intensity varies rapidly across phase space, and the finite resolution causes a nontrivial migration of events.
Eq.~(\ref{eq:f_reso}) is designed to correct precisely this local migration effect.

In the  amplitude analysis, the resolution correction is incorporated through event weights.
For a data sample containing a specific number of  reconstructed events ($N_{\mathrm{evt}}$), each event at phase-space point $x_k$ is assigned the factor
$f_{\mathrm{reso}}(x_k)$ computed from Eq.~(\ref{eq:f_reso}).
If $S_k$ denotes an event-wise contribution to the fit, such as a likelihood weight or another event-level quantity, the corrected sum is written as
\begin{equation}
S = \sum_{k=1}^{N_{\mathrm{evt}}} S_k\, f_{\mathrm{corr}}(x_k).
\end{equation}

\begin{figure}[htbp]
\centering
\includegraphics[width=0.39\textwidth]{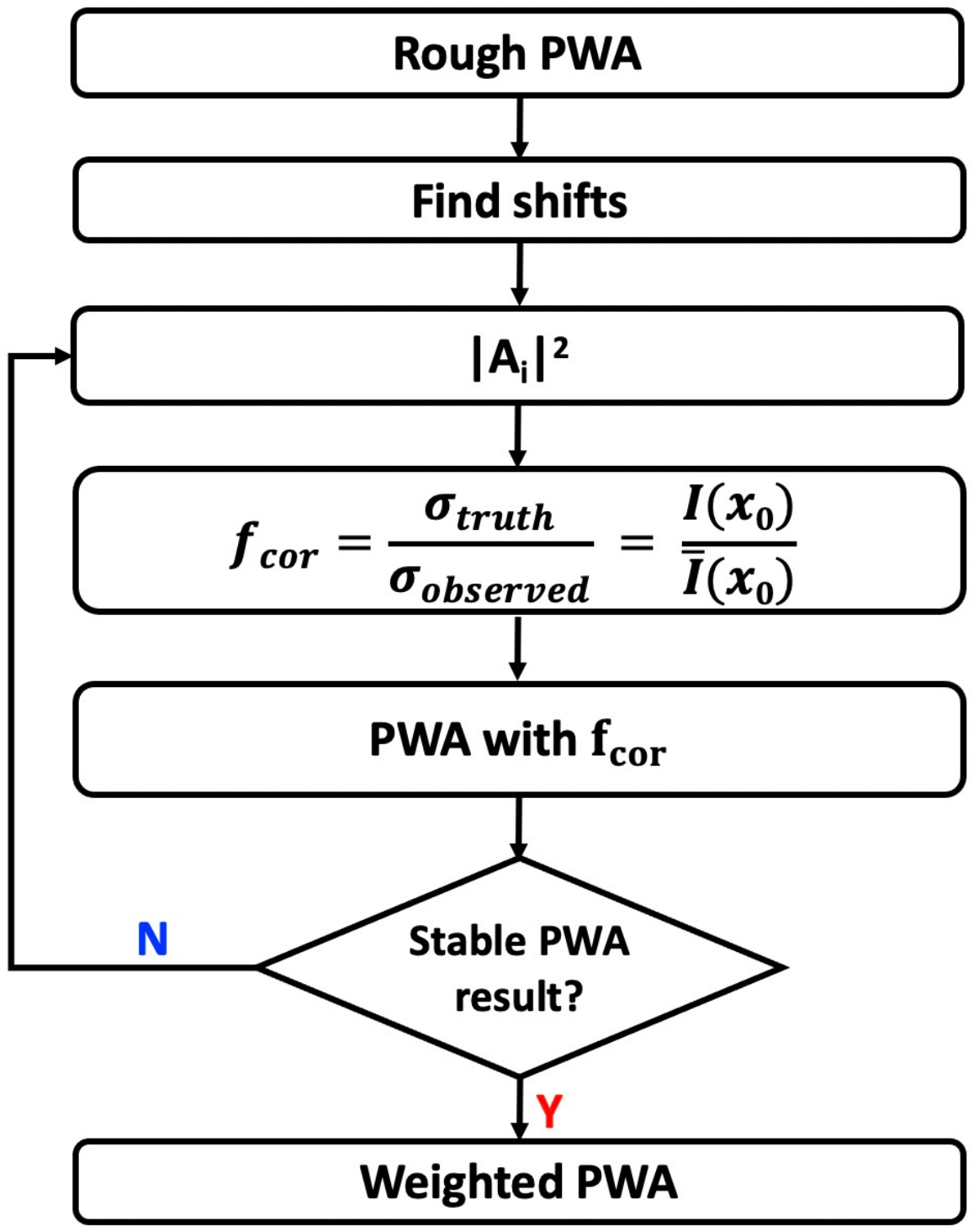}
\caption{Flowchart for this reweighting method.} 
\label{flowchart}
\end{figure}

Since the correction factor depends on the amplitude model itself, the method is intrinsically iterative.
We begin with an initial amplitude model $A^{(0)}(x)$ obtained from a standard fit to the reconstructed data without resolution correction.
Using this model, the factors $f_{\mathrm{corr}}(x_k)$ are calculated for all data events, and a new fit is then performed with the corrected event weights.
This yields an updated model, from which revised correction factors are obtained.
The procedure is repeated until the fit parameters of interest, such as the masses and widths of intermediate resonances, become stable within a predefined tolerance. And the 
 flowchart for this method is illustrated  in Fig.~\ref{flowchart}.

\section{Example and Performance}

A compelling demonstration of the weighting method's validity is its application to the analysis of the decay \(J/\psi \rightarrow \Xi^{-} \bar{\Xi}^{+}\pi^{0}\).   
A. MC sample of \(J/\psi \rightarrow \Xi^{-} \bar{\Xi}^{+}\pi^{0}\), including the intermediate state of $\Xi(1530)$ and the the phase space contribution, 
   is generated to investigate this method in which the mass and width of them
are from the world average values~\cite{pdg2026}.   And the relevant kinematic variables could be be taken as
\begin{equation}
x \equiv \left(m^2_{\Xi\pi^0},\, m^2_{\bar{\Xi}\pi^0},\, m^2_{\Xi\bar{\Xi}}\right),
\end{equation}
with only two independent components required in the Dalitz-plane distance measure.

\begin{figure}[htbp]
\centering
\includegraphics[width=0.39\textwidth]{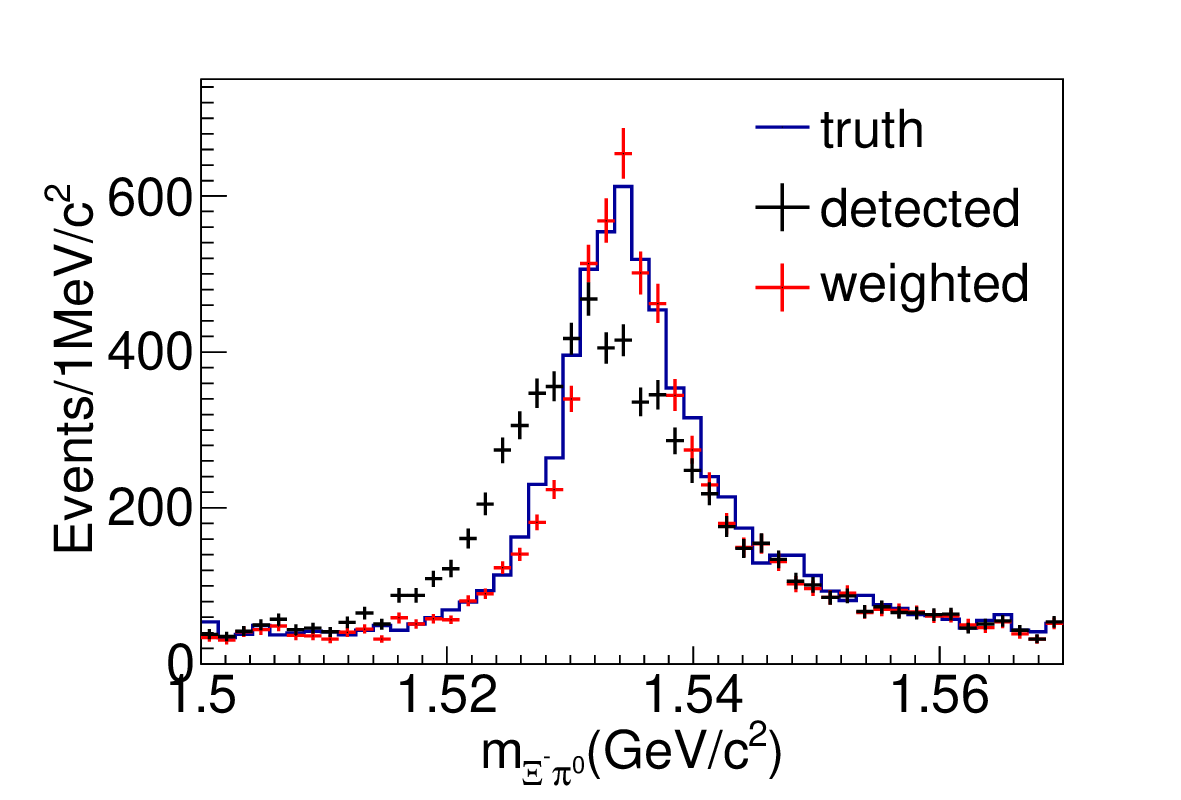}
\caption{The  distribution of $m_{\Xi^{-}\pi^0}$ in the vicinity of the $\Xi(1530)$ peak. The histogram represents  the truth distribution, black crosses denote the detected distribution and the red crosses  are for the weighted distribution.} 
\label{mxipi}
\end{figure}

 \begin{figure}[htbp]
\centering
\includegraphics[width=0.39\textwidth]{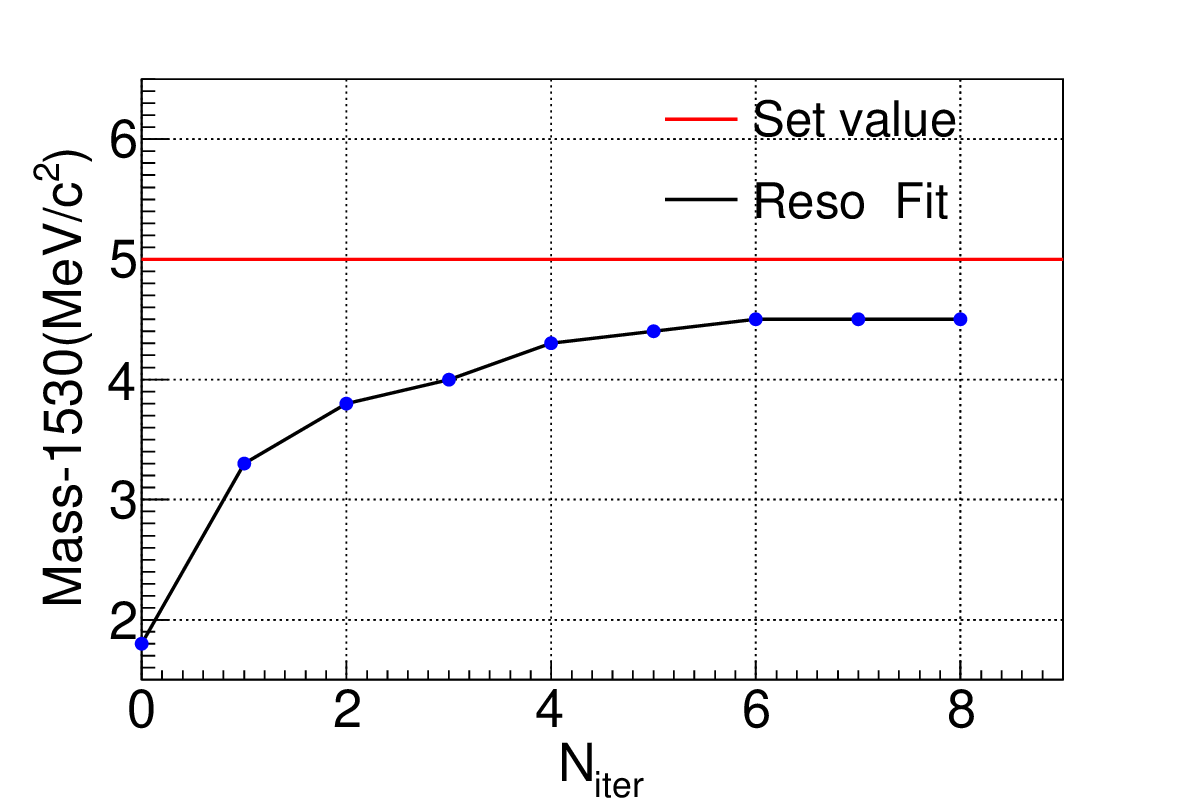}
\caption{
Fitted mass of the $\Xi(1530)$ from the PWA as a function of the weighting iteration. The red line indicates the input (set) value ($m=1535$ MeV), and the dots show the PWA output after the $n$th weighting iteration. The point at 
$N_{\rm iter}=0$ corresponds to the result without weighting. } 
\label{mass}
\end{figure}

\begin{figure}[htbp]
\centering
\includegraphics[width=0.39\textwidth]{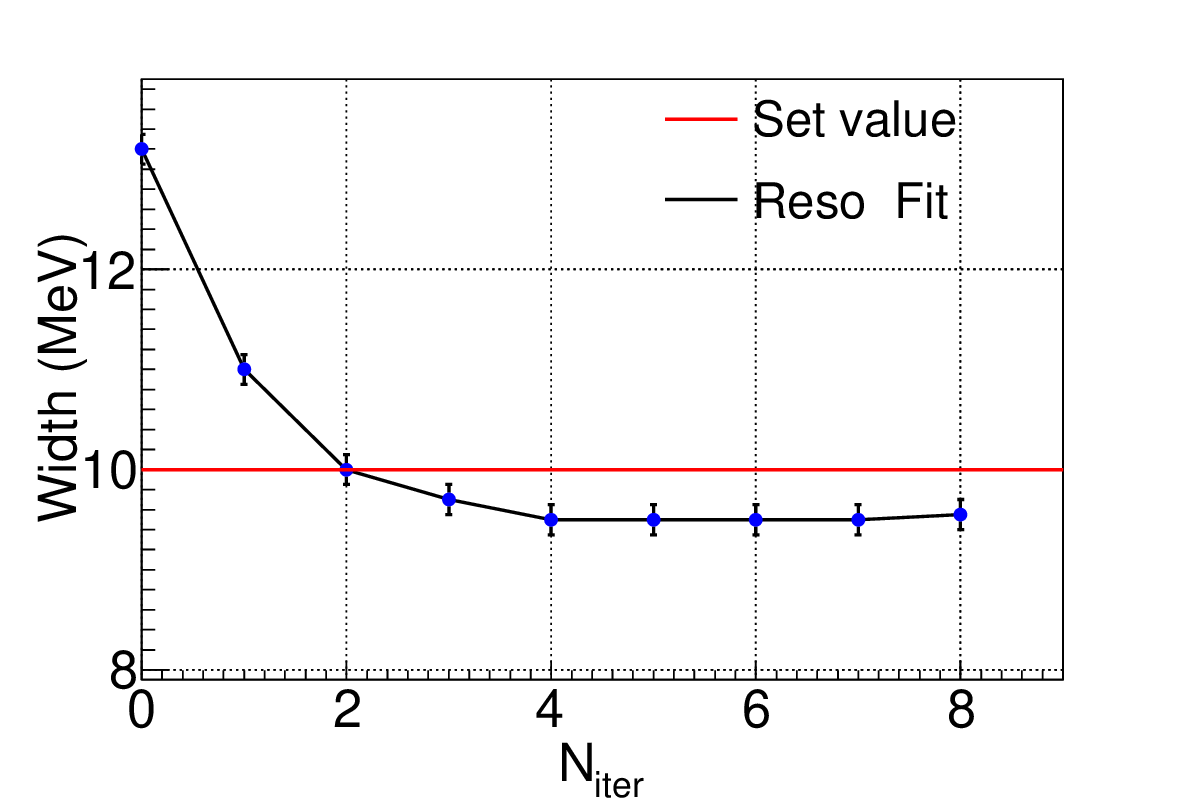}
\caption{Fitted width of the $\Xi(1530)$ from the PWA as a function of the weighting iteration. The red line indicates the input (set) value ($\Gamma=10$ MeV), and the dots show the PWA output after the $n$th weighting iteration. The point at 
$N_{\rm iter}=0$ corresponds to the result without weighting.} 
\label{width}
\end{figure}

 \begin{figure}[htbp]
\centering
\includegraphics[width=0.39\textwidth]{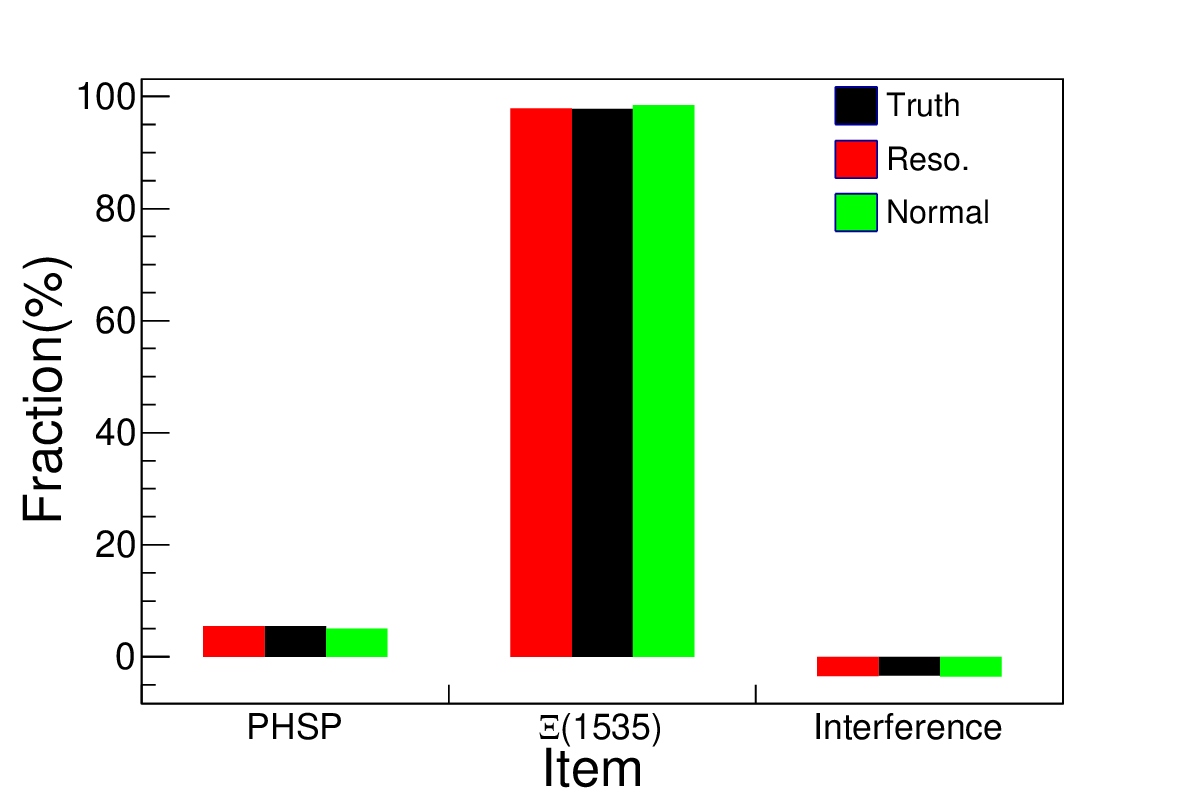}
\caption{
The comparisons of the fractions for  $\Xi(1530)$, the phase space contribution and the interference between them at the truth level, with and without weighting. } 
\label{fraction}
\end{figure}

As shown in Fig.~\ref{mxipi}, the detected distribution of $m_{\Xi\pi^0}$ vicinity of  $\Xi(1530)$ peak (black crosses) deviates from the truth (blue histogram) clearly.
Therefore, without resolution treatment, the PWA fit can be misled by the smeared data. The performance is then quantitatively assessed by comparing PWA fits performed on two  cases:  one without resolution effects and one where the weighting method has been applied to account for resolution.  As illustrated by the weighted distribution (red error bars) in Fig.~\ref{mxipi}, the comparison  confirms the validity of the method,  indicating improved agreement between the fitted data distribution and the truth-level distribution.   
We  also checked the fitted mass (\(m\)) and width (\(\Gamma\)) as illustrated in  Fig.~\ref{mass} and
 Fig.~\ref{width}, respectively,  where the output mass and width of $\Xi^*$ in the PWA results get stable and close to the truth value step by step. 
 Additionally,  as shown in Fig.~\ref{fraction}, the weighting method is also found to reduce the bias in the fitted fractions of individual components.

\begin{figure*}[htbp]
\centering
\includegraphics[width=0.32\textwidth]{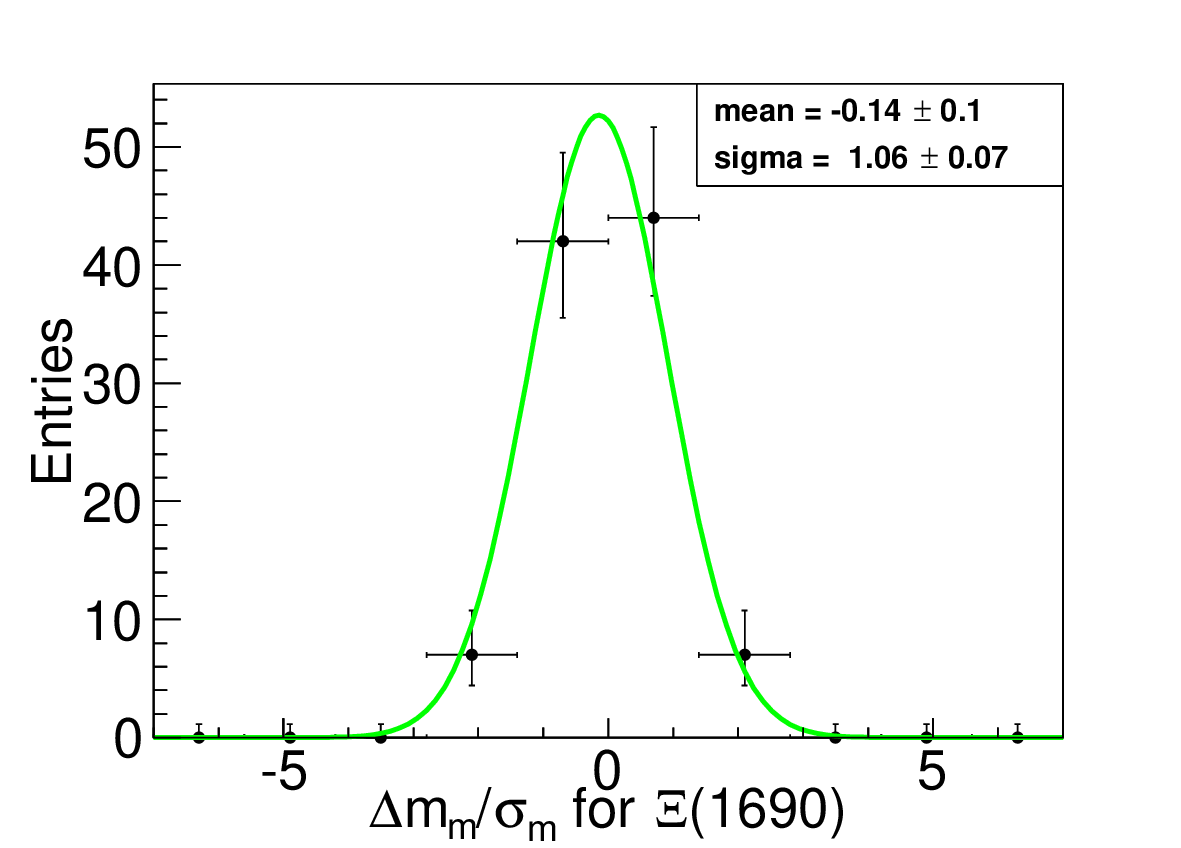}
\includegraphics[width=0.32\textwidth]{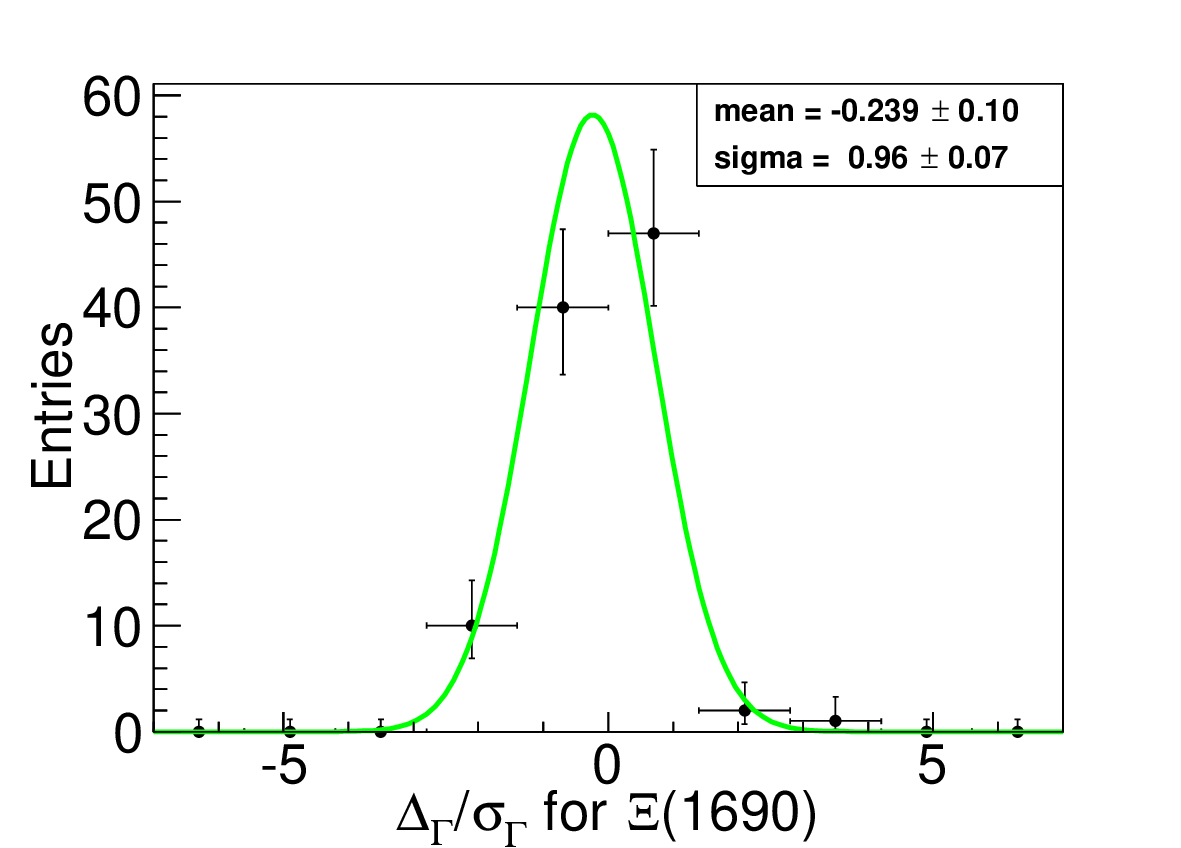}
\includegraphics[width=0.32\textwidth]{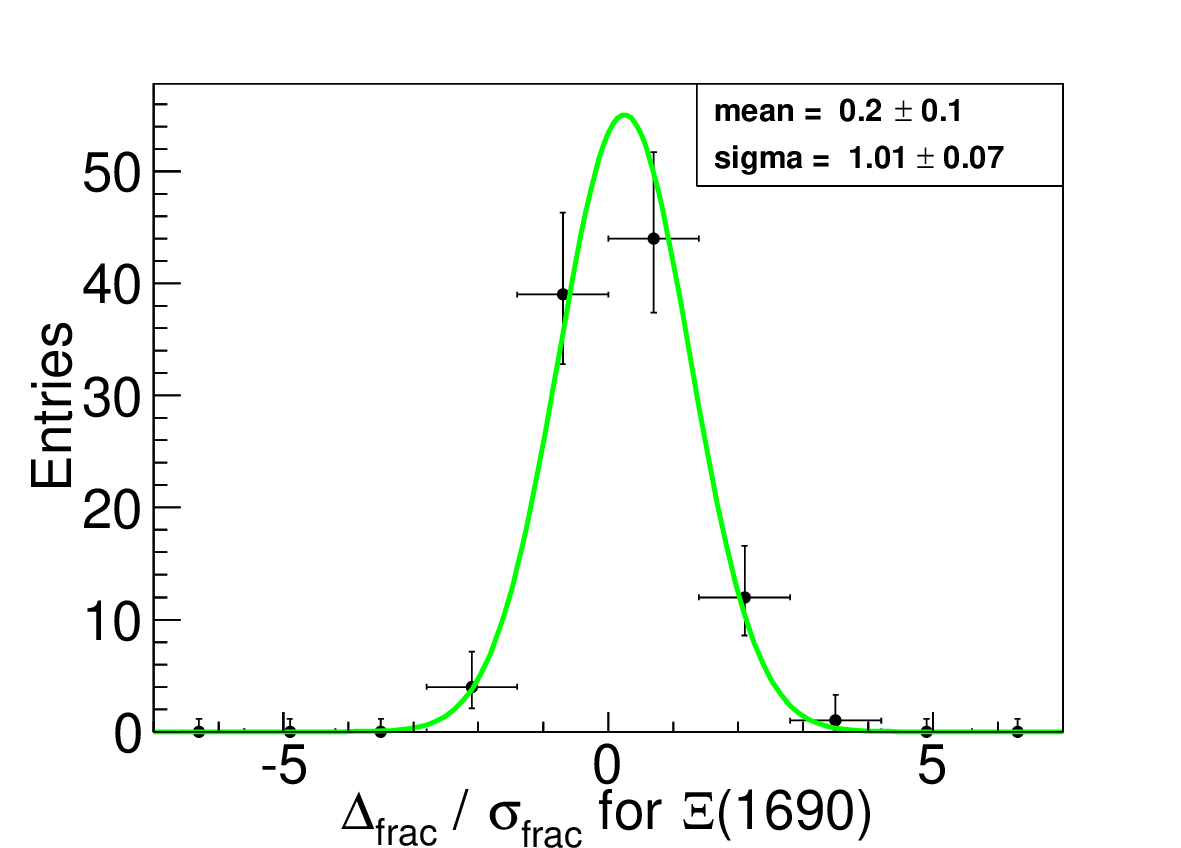}

\caption {The comparisons of  each fitted parameter for  $\Xi(1690)$  obtained from a series of the weighted PWA, which could be described with a standard Gaussian function.} 
\label{mandw1}
\end{figure*}

\begin{figure*}[htbp]
\centering

\includegraphics[width=0.32\textwidth]{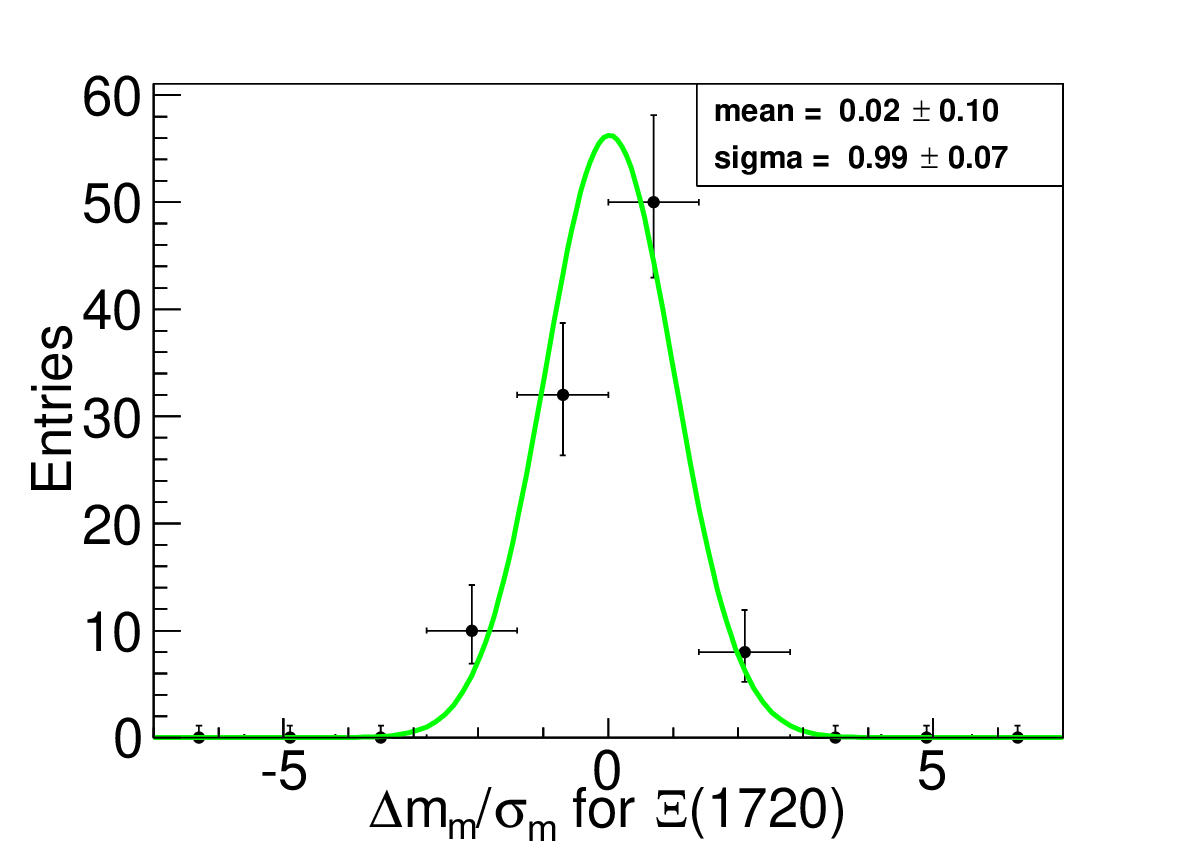}
\includegraphics[width=0.32\textwidth]{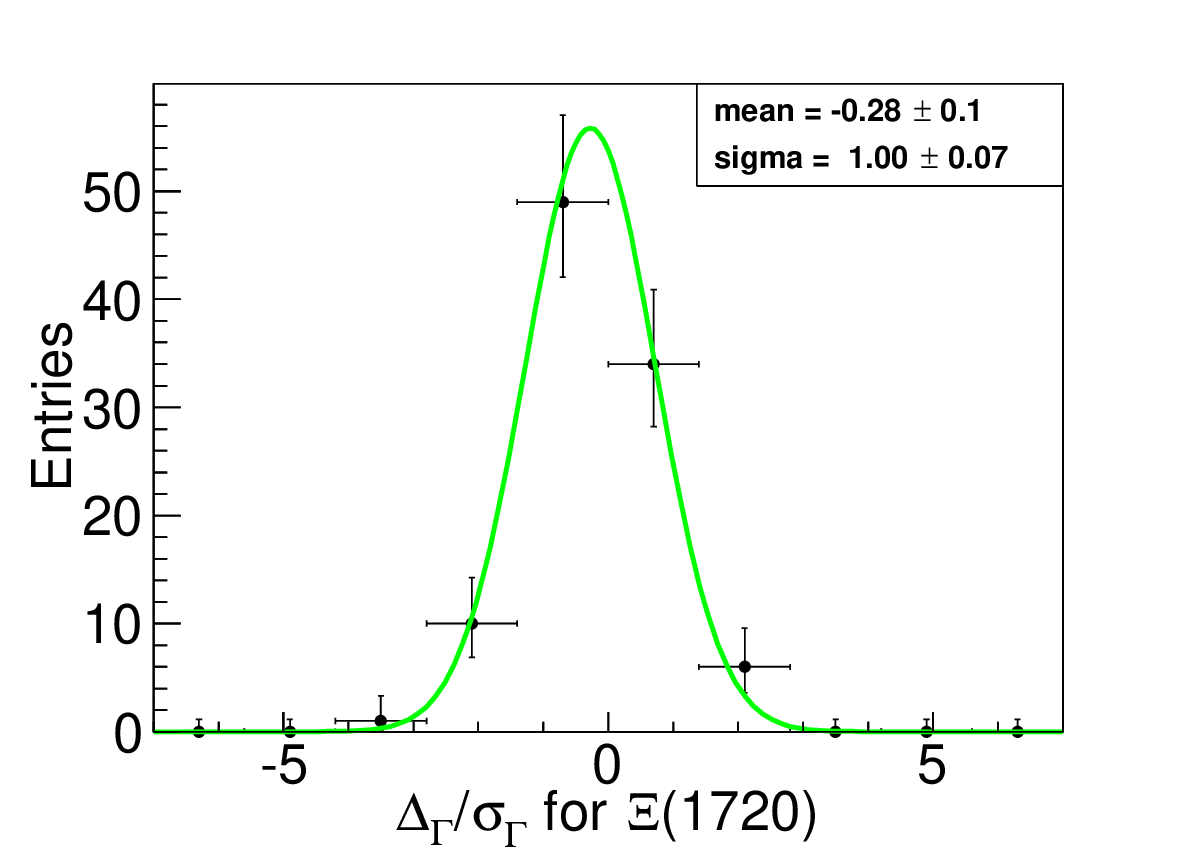}
\includegraphics[width=0.32\textwidth]{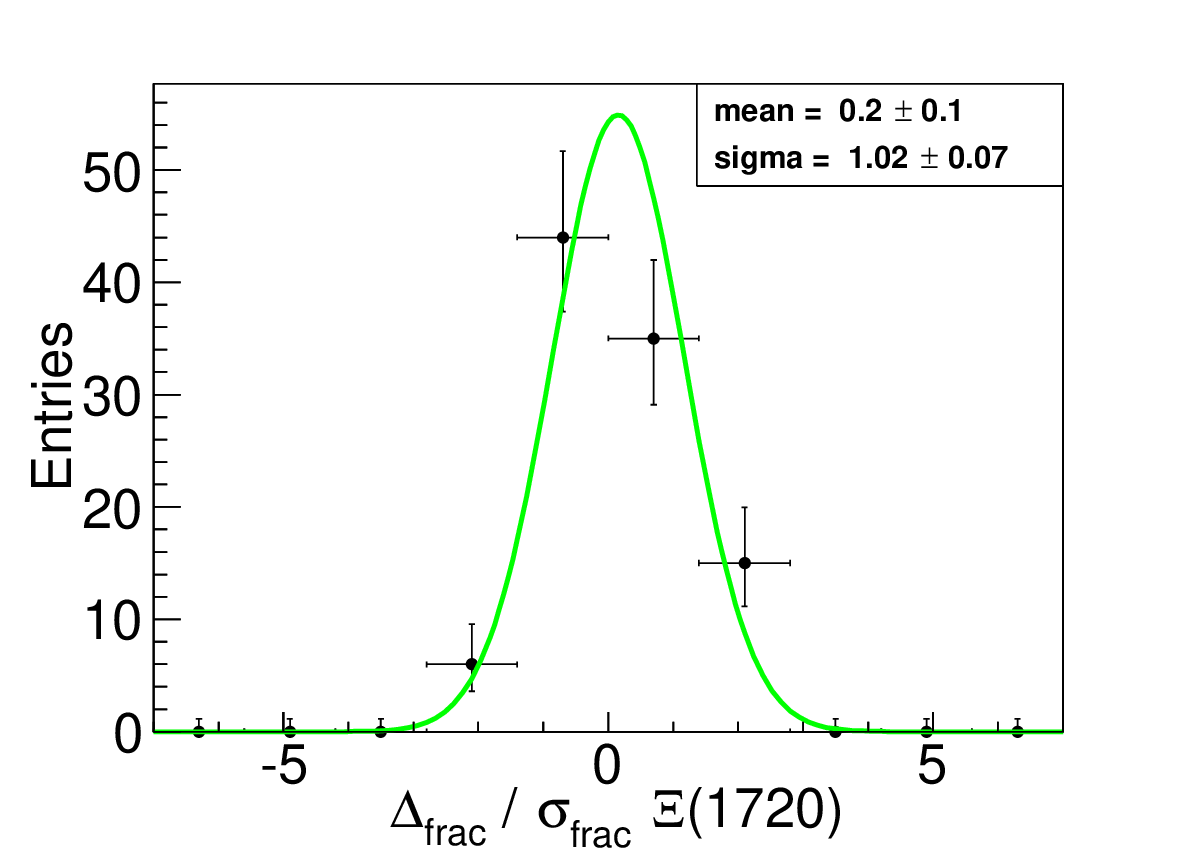}

\caption{The pull distributions of each fitted parameter for  $\Xi(1720)$  obtained from a series of the weighted PWA, which could be described with a standard Gaussian function.} 
\label{mandw2}
\end{figure*}

To further study the weighting method,  we generated an MC sample of $J/\psi\to K^-\Lambda\bar{\Xi}^+$, containing  $\Xi(1690)$, $\Xi(1720)$ and non-resonant contributions, together with their mutual interference, in which the masses and widths of the intermediate states are from PWA of $J/\psi\to K^-\Sigma^0\bar{\Xi}^+$~\cite{BESIII:2026jcv}. After accounting for the mass resolution, we  checked the fitted returns such as the mass and width  of the  intermediate states  as well as the fitted fractions and the comparisons are presented in Fig.~\ref{mandw1} and Fig.~\ref{mandw2},  respectively. 
After applying the event weight correction,  we found that the input and output values for mass and width of   both $\Xi(1690)$, $\Xi(1720)$ are follows the Gaussian function, 
these biases are drastically reduced. The data indicates that the deviation of fitted parameters from their true generated values becomes much smaller with the weighting method applied. In this sense, this method successfully deconvolves the detector effects, allowing the theoretical model to describe the true physics rather than the detector-smeared data.

The method's robustness is highlighted by several key features observed in this and other tests. First, it requires minimal computational overhead. Once the weight tables are generated from a sufficiently large simulated sample, they can be reused in the fit. Second, it is effective for both wide and very narrow resonances. For states whose natural width is significantly larger than the resolution, the method often works well without iterative refinement. For narrow states, it is particularly beneficial, enabling precise measurements of mass and width and a more reliable determination of spin-parity, which is crucial for discovering new resonances.

Furthermore, the technique is not limited to mass resolution. As noted, it can be generalized to correct for momentum resolution and angular resolution. This makes it applicable to analyses that are sensitive to specific momentum or angular distributions, thereby extending the range of physics processes amenable to precise amplitude analysis.  

\section{Conclusion}

In this work, we present a practical weighting method for incorporating detector mass-resolution effects into amplitude analysis. The central idea is to replace an explicit multidimensional convolution of the theoretical intensity with the detector response by a local, simulation-driven average over truth-level phase-space points associated with each observed event. In this way, detector smearing can be included directly in the likelihood evaluation without the prohibitive cost of full detector simulation at every fit iteration.
The method is especially valuable for analyses involving narrow intermediate resonances, where finite resolution can distort line shapes, shift fitted masses and widths, and even generate fake structures in the spectrum. By constructing event-by-event weights from nearby truth–reconstruction correspondences in MC samples, the approach naturally accounts for phase-space-dependent migration and can be extended beyond invariant masses to momentum and angular variables that are essential for spin-parity studies.

Applications to the decays  $J/\psi \to \Xi^{-} \bar{\Xi}^{+}\pi^0$  and $J/\psi\to K^-\Lambda\bar{\Xi}^+$  
  demonstrate that the method significantly reduces biases in resonance parameters and fit fractions, leading to a more relilable description of the underlying dynamics. At the same time, the procedure remains computationally efficient and flexible enough for iterative use in realistic partial-wave analyses.
  
Although  some model dependence remains through the assumed amplitude, the iterative correction scheme helps mitigate these limitations. Overall, this weighting method provides a reliable and efficient tool for improving the precision and robustness of amplitude analyses in modern high precision hadron spectroscopy experiments, such as BESIII~\cite{BESIII:2009fln} and LHCb~\cite{LHCb:2014se,LHCb:2008vvz}.


\section{Acknowledgments}
 This work is supported in part by National Key R\&D Program of China under Contracts No. 2025YFA1613900; National Natural Science Foundation of China (NSFC) under Contracts Nos. 12225509, 12475089; Guangdong Basic and Applied Basic Research Foundation 2024A1515012416.




\end{document}